# Sagittarius: The Nearest Dwarf Galaxy

Rodrigo A. Ibata[1,2], Gerard F. Gilmore[1] and Michael J. Irwin[3]

1 *Institute of Astronomy, Madingley Road, Cambridge CB3 0HA*
2 *Department of Geophysics and Astronomy, 2219 Main Mall, University of British Columbia, Vancouver, Canada V6T 1Z4 (present address)*
3 *Royal Greenwich Observatory, Madingley Road, Cambridge CB3 0EZ*

**ABSTRACT**
We have discovered a new Galactic satellite galaxy in the constellation of Sagittarius. The Sagittarius dwarf galaxy subtends an angle of $\sim 10$ degrees on the sky, lies at a distance of 24 kpc and is comparable in size and luminosity to the largest dwarf spheroidal, Fornax. The new galaxy has many features in common with the other eight Galactic dwarf spheroidal systems: including an extended low density spatial structure; a well populated red horizontal branch with a blue extension; and a substantial carbon star population. In terms of stellar populations it most closely resembles the Fornax dwarf, having a strong intermediate age stellar component and evidence of a metallicity spread. Sagittarius is the nearest galaxy known and currently lies only $\sim 16$ kpc from the centre of the Milky Way. Isodensity maps show it to be markedly elongated along a direction pointing towards the Galactic centre and suggest that it has been tidally distorted. The close proximity to the Galactic centre, the morphological appearance and the radial velocity of 140 km/s indicate that this system must have undergone at most very few close orbital encounters with the Milky Way. It is currently undergoing strong tidal disruption prior to being integrated into the Galaxy. We find that at least some of the four globular clusters, M54, Arp 2, Ter 7 and Ter 8, are associated with the Sagittarius dwarf galaxy, and will probably share the fate of their progenitor. The Sagittarius dwarf galaxy was found serendipitously using a combination of UK Schmidt Telescope sky survey plates, the APM automatic plate measuring facility and the Anglo Australian Telescope multifibre spectrograph, AUTOFIB.

## 1 INTRODUCTION

Galaxies of the dwarf spheroidal class (dSph) are among the smallest and faintest known. Eleven such objects have been discovered so far near the local group spirals: three are companions to M31, and a further eight surround the Milky Way. The first seven of the Milky Way satellite galaxies were discovered by visual inspection of photographic plates; whilst the eighth was found from machine measurements of a photographic plate (Irwin et al. 1990). Their surface brightness is so low that all have been discovered from the concentration of resolved stars over and above that of the foreground component of our own galaxy. These dwarfs, also known as Sculptor-type systems, or dwarf elliptical galaxies, range in diameter from 1-5 kpc, subtend angles of up to several degrees on the sky, have absolute magnitudes in the range from -12 to -8 and central surface brightnesses ranging from 22 mag/arcsec$^2$ down to 27 mag/arcsec$^2$. Taken together with the dwarf irregulars these galaxies are by far the most numerous members of the Local Group.

In most galaxy formation scenarios, primordial density fluctuations grow rapidly after the inflationary epoch to become centres for the formation of structure in the Universe. In so-called 'bottom-up' models, large scale structure is built from the merging of smaller entities. In this picture, typical $1\sigma$ primordial fluctuations can give rise to a dwarf galaxy sized perturbation, which evolves either into a dwarf galaxy as seen at the present epoch, or is incorporated into a more massive galaxy (see eg. Silk et al. 1987). It is interesting in this context that dark matter is a significant constituent of dwarf galaxies. This fact is deduced from the large discrepancies between the mass found by applying the Virial theorem to the observed velocity dispersion of tracer stars compared to that mass inferred from their integrated luminosity (*cf* Hargreaves et al. 1994 and references therein). If this scenario is correct, dwarf spheroidals, despite their unassuming appearance, may hold important clues regarding the reasons for and conditions of the initial collapse, subsequent evolution and final structure of the Milky Way and other large galaxies.

Regions of sky at relatively high Galactic latitude, $|b| \gtrsim 20°$, have been thoroughly surveyed for Milky Way satellite galaxies both by eye and by using a combination of Schmidt sky survey plates and the automatic plate measuring (APM) facility in Cambridge (Irwin et al. 1990). Apart from Sextans, no new dwarf spheroidal galaxies were found in the automated survey, in spite of its much increased sensitivity compared to previous efforts. Unfortunately, general searches at lower Galactic latitudes are compromised by a combination of the high foreground stellar density, the strong gradient in stellar density toward the Galactic plane and by patchy Galactic extinction. However, with the increasing availability of high throughput multi-object spectrographs it is possible directly to search for comoving groups of stars close to the Galactic plane and, often, as in this case, as the byproduct of a different investigation.

The study of moving groups of stars (ie. stars with a common distance and velocity) has a long and varied history (Eggen, 1971). Such moving groups may represent the relics of recently disrupted clusters or tidally stripped satellite galaxies. The unambiguous detection of such a system would represent an ideal opportunity to study the processes at work in the disintegration of stellar systems. However, although several authors have reported finding comoving groups of stars (Arnold & Gilmore 1992, Côté et al. 1993, Majewski 1992, Rodgers et al. 1990, Ratnatunga 1988, Sommer-Larsen & Christensen 1987), the numbers of such stars are generally very small (∼10) and the interpretation is often subject to considerable uncertainty. We report in this article on the detection of a large extended comoving group close to the Galactic centre and show that it is most probably part of a previously unknown dwarf satellite galaxy. Following convention, we propose to call the new system the Sagittarius dwarf galaxy.

# 2   DISCOVERY

As part of an investigation into stellar abundances and kinematics in the outer regions of the Galactic Bulge, eighteen regions were initially selected for study from the reddening maps of Burstein & Heiles (1982), along lines-of-sight known to have low extinction. Candidate K and M giant stars were chosen for spectroscopic followup on the basis of APM measures of UKST B$_J$ and R plates. The spectroscopic data were collected over three observing runs using the 3.9m Anglo-Australian Telescope (AAT) equipped with the multifibre system AUTOFIB. This spectrograph allows simultaneous study of some 64 objects, of which 50 are typically targets of direct interest, with the remaining fibres being used for sky estimation (Wyse & Gilmore 1992). A total of forty fibre fields chosen from six of the eighteen selected regions was observed in this way and reduced in the standard manner (*cf* Ibata & Gilmore 1995). Radial velocities for all the target stars, accurate to $9\,\mathrm{km\,s^{-1}}$, were then derived by cross-correlating each target spectrum with radial velocity standard star templates.

# 3   PROPERTIES

## 3.1   Velocity Structure

Figure 1 shows the comparison between the observed Galactocentric radial velocity distributions and that predicted by our Galaxy model in the six fields where spectroscopic data were obtained. The model contains reasonably well established descriptions of the Galactic disk, thick disk and halo, while the description of the Galactic bulge component has been taken from our recent self-consistent gravitational model of the inner Galaxy (Ibata & Gilmore 1995a,1995b). The fits shown in figure 1 are acceptable (in the sense that the probability that they are incorrect is less than 95%) only if one excludes data from the region near $\approx 180\,\mathrm{km\,s^{-1}}$ in three fields at $\ell = 5°, b = -12°$, $\ell = 5°, b = -15°$ and $\ell = 5°, b = -20°$, where the model fails to account for a low dispersion group of more than 100 stars. This unexpected feature is perhaps most striking in the velocity distribution of M giants. Of the 42 M giants that were found in the entire stellar sample, 36 originated from the three fields at $\ell = 5°$; their distribution is shown in figure 2. These M giants are obviously not drawn randomly from the stellar sample displayed in figure 1.

The low dispersion structure cannot be associated with Galactic thick disk, halo, bulge or distant old disk populations if canonical models are correct; furthermore the feature has too large a Galactocentric velocity to be associated with the local thin disk (whose Galactocentric velocity in these fields we have shown, *cf* Ibata & Gilmore (1995b), to be $\approx 20\,\mathrm{km\,s^{-1}}$, as is expected).

We therefore introduce a further term into our kinematic Galaxy model to account for this feature, and assume, for simplicity, that the excess stars are distributed in a gaussian manner. From 'bootstrapping' maximum likelihood fits to the stellar velocity distribution, we find that the systemic heliocentric radial velocity of the excess stars is $+140 \pm 2\,\mathrm{km\,s^{-1}}$ ($172\,\mathrm{km\,s^{-1}}$ in a Galacto-centric system) with a velocity dispersion, corrected for measuring errors, of ∼10 km/s; these fits are shown in figure 3. There is no significant radial velocity difference, < 5 km/s, between the mean radial velocity of the excess stars at $b = -12^\circ$ and those at $-20^\circ$.

Figure 1. Comparison between the expected velocity distribution and the data. The lines of sight are: (a) $\ell = -25^\circ$, $b = -12^\circ$, (b) $\ell = -15^\circ$, $b = -12^\circ$, (c) $\ell = -5^\circ$, $b = -12^\circ$, (d) $\ell = 5^\circ$, $b = -12^\circ$, (e) $\ell = 5^\circ$, $b = -15^\circ$ and (f) $\ell = 5^\circ$, $b = -20^\circ$. The unknown feature is observed in (d), (e) and (f), centred at a velocity of $172\,\mathrm{km\,s^{-1}}$.

Figure 4 shows the relation between colour and velocity of the stars observed in the $\ell = 5^\circ$ fields. The feature is well-populated at very red colours and extends out to $(B_J - R) = 3.2$, in contrast to the bulk of the (bulge) stars, which are concentrated towards the blue end of the diagram. This colour dependency of the kinematic feature showed that a detailed investigation into the colour-magnitude structure of this region of sky was required.

## 3.2 Colour-Magnitude Relation

To further investigate the unknown kinematic feature, we used the APM facility (Kibblewhite et al. 1984) to measure $5.3^\circ \times 5.3^\circ$ areas of UKST survey fields that straddle the region of sky over which the feature had been found. A deep $2.5^\circ \times 2.5^\circ$ photographic colour-magnitude diagram (CMD), calibrated with the CCD photometry described in Ibata & Gilmore (1995a), and centred at $18^h 51.9^m$, $-30°33'$ (B1950) ($\ell = 5.5°$, $b = -14.1°$) was constructed (figure 5a) from these scans.

Figure 2. The galactocentric radial velocity distribution of stars classified as M giants in three fields at $\ell = 5°$, $b = -12°$, $-15°$, $-20°$.

The expected CMD, according to our Galaxy model, along the same direction, over the same solid angle and with the reddening value discussed in section 3.5, is shown in figure 6. Unfortunately, the constraints on Galactic components near the Galactic centre are currently insufficient to allow a direct quantitative comparison to figure 5a, but we know figure 6 to be qualitatively correct, since it reproduces the main features of deep CCD CMDs (cf eg Harding & Morrison 1993). The luminosity information in the model is quantitatively correct to R ≈ 18, as we demonstrate in figure 7 below. In detail, the CMD in figure 6 is a complicated mixture of Galactic components at every point on its surface; nevertheless, it is easy to discern the following characteristics. Bulge giants dominate the red, vertical feature visible from R ≈ 17, $(B_J - R) \approx 1.0$ to R ≈ 14, $(B_J - R) \approx 1.4$, while the blue feature from R ≈ 17, $(B_J - R) \approx 0.7$ to R ≈ 14, $(B_J - R) \approx 0.7$ is primarily due to foreground disk dwarfs. Disk dwarfs also dominate the less populated area in the region of R ≈ 18, $(B_J - R) \approx 1.6$. All Galactic components are present in large numbers in the dense region near R ≈ 18, $(B_J - R) \approx 0.9$.

The unexpected aspects of figure 5a are therefore the large excess of stars at R = 17.7, $(B_J - R) = 1.4$ with possible traces of a long blueward extension to R = 18.0, $(B_J - R) = 0$; and the excess in the form of a steeply rising arm extending to R ∼ 14, $(B_J - R) \sim 3.0$. The reddest and brightest stars in this figure are seen clearly also in figure 4 above, where it is apparent they extend in colour to $(B_J - R) \sim 3.4$. Figure 5b shows an observed comparison CMD, constructed from photometry from the same photographic plate as figure 5a, but centred on $\ell = 8.4$, $b = -13.5$. Though the reddening in this region is almost identical to that at $\ell = 5.5°$, $b = -14.1°$ (*cf* section 3.5), only expected Galactic features (similar to figure 6) are visible. We have generated $2.5° \times 2.5°$ CMDs down to R = 19 in several locations on five UKST fields 396,397,458,459,460 that lie near $\ell = 5°$, $b = -15°$, and on UKST fields 336 and 517 which lie on the other side of the bulge minor axis at $\ell = -5°$, $b = -15°$, and $\ell = -5°$, $b = 15°$ respectively. We have also obtained $1° \times 1°$ CMDs down to R = 17 at the positions of the eighteen selected regions which lie near the intersections of $\ell = 25°, 15°, 5°, -5°, -15°, -25°$ and $b = -12°, -15°, -20°$. The unusual feature visible in figure 5a has only been found in UKST fields 458 and 459.

### 3.3 Luminosity Function

We construct a luminosity function (LF) for the unexpected colour-magnitude population described in section 3.2. This is done empirically, subtracting the luminosity information in figure 5b from that in figure 5a to give the LF in figure 8, and also using the Galaxy model, subtracting the luminosity information in figure 6 from that in figure 5a to give the LF displayed in figure 9. The precise morphology of the LF in figure 9 is dependent on uncertainties in the Galaxy model parameters which are very difficult to quantify; however, the similarity between these plots is at least reassuring. We therefore adopt the lower line in figure 8 as the best estimate of the LF of the feature. (This exercise demonstrates the uncertainties in subtracting a weak feature from a large contaminating population).

### 3.4 Magnitude and Colour Spread of the Unexpected Features

The width of the 'bump' in the LF shown in figure 8 gives an upper limit to the magnitude spread of the excess population near R = 17.7. From a maximum likelihood gaussian fit to that 'bump' we find a spread of $\sigma = 0.19^m$; however, the *rms* photometric error estimated in Ibata & Gilmore (1995a) was also found to be ≈ $0.2^m$, so this result is consistent with the population lying in a narrow distance interval along the line of sight, and having a narrow range in intrinsic luminosity.

In figure 10 we show that the population near the end of the arm that rises from R = 17.7, $(B_J - R) = 1.4$ to R = 14, $(B_J - R) = 3.0$ has a colour spread of ≈ $0.2^m$.

### 3.5 Reddening

The maps of Burstein & Heiles (1982) show low reddening over the regions investigated;

Figure 3. Comparison between the observed velocity distribution and that expected from the standard Galaxy model plus a gaussian component of variable mean, dispersion and normalisation which is included so as to account for the feature near $172\,\mathrm{km\,s^{-1}}$. The lines of sight are the same as in figure 1.

from these we find the extinction in $(B - V)$ in field **458** to be $0.14^{\mathrm{m}}$. **According to these maps, the reddening is uniform to within** $0.03^{\mathrm{m}}$ **over the area of the scan. Furthermore, we have shown that the excess of stars at R = 17.7 and the 'arm' from R = 17.7, $(B_J - R)$ = 1.4 to R = 14, $(B_J - R)$ = 3.0 in figure 5a are concentrated over a colour range of equal width to the photometric error estimate, therefore the colour-magnitude data are also consistent with** there being negligible reddening variation along that line of sight. That is, the reddening is predominately foreground. Other estimates of the reddening in this region have been obtained from studies of the properties of the globular cluster M54, which lies at $\ell = 5.6°$, $b = -14.1°$, near the centre of the scan from which figure 5a was constructed. From the mean colour of RRLyraes in M54, **Harris (1975) finds** $E_{B-V} = 0.17$, **in good** agreement with the work of Harris & van den Bergh (1974) who find $E_{B-V} = 0.15$ from the integrated colour of the cluster. The corresponding extinction is $A_V = 0.4^m$. (The reddening vector in the $(B_J - R)$, R colour system is superimposed on figure 6).

Figure 4. The heliocentric radial velocity – colour distribution of the sample of stars observed at $\ell = 5°$, $b = -12°$, $-15°$, $-20°$. The moving group of stars is centred at a velocity of 140 km/s. (Colours have not been corrected for reddening).

To investigate the reddening variation over the $2.5° \times 2.5°$ field from which figure 5a was created, we divided that field into 16 subfields, each of area $0.63° \times 0.63°$. The magnitude position of the maximum of the excess population in these subfields varies by less than $0.1^m$ over the area of the $2.5° \times 2.5°$ scan. Expected variations in photometry over the whole area of a UKST plate are $\approx 0.05^m$ (Maddox 1988). Thus reddening variations are typically smaller than 0.04 in $(B - V)$ in this field, consistent with the reddening maps of Burstein & Heiles (1982).

### 3.6  Spatial Structure

The presence of the well populated excess at R = 17.7, $(B_J - R) = 1.4$ provides a convenient way to determine the spatial extent and luminosity of the feature against the contamination from the expected Galactic stars. However, the large gradient due to expected Galactic stars precludes direct inspection of its spatial extent. A simple solution is to produce an isopleth (isodensity) map containing images selected to be in the range $17.4 < R < 18.0$ and subtract off from it an equivalent normalised isopleth map containing images at different magnitudes. The overdensity of the features in this apparent magnitude range then reveals any underlying structure. Our preliminary mapping of four UKST fields 396,397,458,459 in this way, shows that the feature extends over fields 458 and 459 but is not apparent in the adjacent fields. In figure 11 we show the distribution of these stars in fields 458 and 459. The feature is elongated on the sky along the direction that points towards the Galactic centre, and appears to have two main subclumps, the centres of which lie at $18^h 51.9^m$, $-30°33'$ and $19^h 03.0^m$, $-30°30'$ (B1950). From previous experience of mapping low surface brightness objects in this way (Irwin & Hatzidimitriou 1993), we estimate that the last isophote would trace $\approx 50\%$ of the whole population, for a dwarf galaxy with an unperturbed density profile. In this case, the enclosed fraction is probably substantially less than this. Future investigations are required to quantify this point further.

From $b \approx -10°$ to $b \approx -20°$ in figure 11, the reddening changes from $E_{B-V} \approx 0.15$ to $E_{B-V} \approx 0.08$, according to the maps of Burstein & Heiles, this is supported by reddening estimates along the lines of sight to the globular clusters listed in table 2. If the reddening is as low as those maps indicate, then figure 11 will be a good representation of the surface density of the unknown feature, since we estimate that it requires a reddening variation of $\delta(B - V) = 0.5$ to alter the counts in figure 11 by one contour level. It should be noted that although the reddening in this region of sky is low and uniform, this is not uncommon at negative latitudes: Harding & Morrison (1993) find a uniform $E_{B-V} \approx 0.07$ in a $2.5° \times 2.5°$ field at $\ell = -10°$, $b = -10°$.

Interestingly, the globular cluster M54 sits at the centre of the denser clump.

### 3.7  Metallicity

Metallicities of the spectroscopically observed survey stars were calculated via an empirical fit of [Fe/H] against $(B - V)$ and MgI'b' equivalent width, which had been obtained from a set of metallicity standard stars. The method is described in detail in Ibata & Gilmore (1995a,1995b), where it is applied to a large sample of Galactic bulge K giants. However, only 8 K giant standard stars with $(B_J - R) > 1.8$ were available for the calibration. The empirical [Fe/H] fits were found to be a strong function of colour, so unfortunately the metallicities of red K giants derived in this way are not very reliable. The external error on a single [Fe/H] measurement is estimated to be $\approx 0.5$dex.

The metallicity – velocity – colour behaviour of the stars in the $\ell = 5°$ region is explored in figure 12. We split the data into two groups below and above a cut at $(B_J - R) = 1.8$; this colour cut was chosen such that $\approx 1/2$ of the stars with Heliocentric radial velocities $> 100\,\mathrm{km\,s^{-1}}$ in the bluer colour group were accounted for with our standard Galaxy model. We further divide each of the colour groups into two subgroups, separating those stars with velocities below and above $100 \, \mathrm{km \, s^{-1}}$. We see that in the bluer colour group, there is no difference between the metallicity distributions of the high and low velocity subgroups (compare figure 12c and figure 12d). However, the redder colour group clearly separates into a bulge population with $v < 100 \, \mathrm{km \, s^{-1}}$, $\overline{\mathrm{[Fe/H]}} \approx -0.5$ and a population with $v > 100 \, \mathrm{km \, s^{-1}}$, $\overline{\mathrm{[Fe/H]}} \approx -1.25$.

Figure 5. Photographic colour-magnitude diagrams (CMD) covering $2.5° \times 2.5°$ of sky. To aid visual inspection, we show the CMDs (a) and (b) with different point widths. The CMD in (a) is centred at $18^h 51.9^m$, $-30°33'$ (B1950) ($\ell = 5.5°$, $b = -14.1°$). Note the dense population at R = 17.7 , $B_J$ − R = 1.4, the blueward extension to $B_J$ − R = 0 and a steeply rising feature extending to R = 14, $B_J$ − R = 3.0, corresponding to the reddest stars seen in Figure 4 above. In contrast, the CMD in (b), from the same UKST plate, but centred at $18^h 54.5^m$, $-27°49.6'$ (B1950) ($\ell = 8.4$, $b = -13.5$), shows only expected features (*cf* figure 6).

We emphasise that calibration limitations with our technique for the coolest stars − we did not expect to see any such stars when beginning our bulge survey! − are such that only the internal differences between figures

Figure 6. Expected colour-magnitude diagram covering $2.5^\circ \times 2.5^\circ$ of sky. The CMD is centred at $18^h 51.9^m$, $-30^\circ 33'$ (B1950). The reddening vector in this colour system is also shown.

Figure 7. A comparison between the expected luminosity function at $\ell = 8.4$, $b = -13.5$ (star symbols) and the observed luminosity function (dots) along the same line of sight (derived from figure 5b). We see clearly that the model reproduces starcounts as a function of apparent magnitude reasonably well down to R $\approx$ 18; below this value, plate incompleteness problems (seen clearly in figure 5) are expected to significantly reduce the number of observed stars. In subsequent plots, the expected LF will be corrected for the incompleteness effect.

Figure 8. The upper line shows the empirical luminosity function of the unknown colour-magnitude feature, created by subtracting figure 5b from figure 5a. The excess population shows up as a large bump near R = 17.7. Counts in this LF will have been overestimated, since figure 5a should contain $\approx 10\%$ more stars than figure 5b, according to the standard Galaxy model. The lower line takes this expected ratio into account. The actual difference in number counts is $\approx 20\%$, the remainder being stars belonging to the population under study.

Figure 9. This shows the luminosity function of the unknown colour-magnitude feature, but this time created by subtracting the luminosity information in the modeled CMD (figure 6) from the observed CMD (figure 5a). Again, the excess population shows up in the form of a large bump near R = 17.7.

12c and 12d, and between figures 12g and 12h, are meaningful. Apparent differences between figures 12c and 12g are not reliable.

To check the metallicity distributions shown in figure 12, the metallicity − velocity data were divided into groups by taking small $(0.2^m)$ colour intervals, thus removing most of the colour dependency of the [Fe/H] fits. A significant difference was found between the metallicity distributions of the bulge and the excess population for all groups with $(B_J - R) > 1.8$, thereby supporting the above result.

### 3.8    Carbon Star Population

Five carbon stars were found in the our spectroscopic survey of the bulge. Of these, one is located at $\ell = -5^\circ$, $b = -12^\circ$, while the other four lie in the $\ell = 5^\circ$ regions. The radial velocities of these stars were calculated by crosscorrelation against radial velocity standard stars. The stars were also crosscorrelated against each other; the internal radial velocity differences thus measured were consistent with the measured 9km/s rms velocity error (*cf* Ibata & Gilmore 1995a).

The spectra of the four stars in the $\ell = 5^\circ$ fields are displayed in figure 13; these stars have heliocentric radial velocities within

Figure 10. In (a), we display the CMD of figure 5a, and show the colour-magnitude region which was selected to give the colour distribution shown in (b). The line through the middle of the box in (a) corresponds to the origin of the abscissa in (b). There is significant contamination from expected Galactic stars on the left hand side of (b), but from fitting a gaussian to the right hand region, we find $\sigma \approx 0.2^{\mathrm{m}}$.

Figure 11. Isopleth maps for UKST fields 458 and 459 constructed from the excess of images present at the horizontal branch magnitude. The lowest contour is 1/8 image arcmin$^{-2}$. Contours increment by 1/2 image arcmin$^{-2}$. The position of the globular cluster M54 is represented by a 'star' symbol.

$20\,\mathrm{km\,s^{-1}}$ (two velocity dispersions) of $140\,\mathrm{km\,s^{-1}}$. The strong $C_2$ absorption feature near $5165\mathring{A}$ can be seen clearly in the spectra. The observed properties of these stars are listed in table 1.

## 4   COMPARISON WITH OTHER STUDIES

### 4.1   Colour-Magnitude Structure

Several unexpected colour-magnitude features have been observed along lines of sight to the Galactic bulge in recent investigations. These include the finding of colour-magnitude substructures (Terndrup 1988, Tyson 1991 which have now been associated with spiral structure by Paczyński et al. (1994). Paczyński et al. also find a concentration of stars near $V \approx 17$, $(V - I) \approx 1.8$, which they interpret to be the bulge red clump at a distance of $8\,\mathrm{kpc}$. Note that this cannot be associated with the excess population observed in figure 5a, unless the extinction in R in that field is $\approx 2.4^{\mathrm{m}}$ (a factor of $\approx 7$) greater that the adopted value.

Table 1. The four carbon stars found in the $\ell = 5°$ fields. The velocities listed are Heliocentric radial velocities.

| Star | $(B_J - R)$ | R | $v(\mathrm{km\,s^{-1}})$ |
|------|-------------|-------|--------------------------|
| 1 | 2.31 | 14.56 | 141 |
| 2 | 2.57 | 14.69 | 136 |
| 3 | 3.01 | 14.33 | 128 |
| 4 | 3.22 | 14.36 | 140 |

Rodgers et al. (1990) obtained a sample of eighteen stars with magnitudes in the range $17 < V < 18$ in a field at $\ell \approx 10°$, $b \approx -22.3°$. They divided the sample into two groups according to their metallicity; the high metallicity group of eight stars with $[\mathrm{Ca/H}] > 0.4$ were found to have a low velocity dispersion ($\sigma = 36\,\mathrm{km\,s^{-1}}$), while the metal poor group displayed $\sigma \approx 100\,\mathrm{km\,s^{-1}}$. The nature of that group remains unclear. However, the metal-rich, low dispersion group presented by Rodgers et al.

Figure 12. We display the metallicity of K giant stars selected from the $\ell = 5^\circ$ fields as follows. (a) shows the Heliocentric radial velocity – metallicity diagram for stars with $(B_J - R) < 1.8$; the corresponding velocity distribution is shown in (b). The standard Galaxy model accounts for $\approx 1/2$ of the stars with Heliocentric radial velocities $> 100\,\mathrm{km\,s^{-1}}$. In (c) and (d) we show the metallicity distribution of these stars below and above a cut at a Heliocentric radial velocity of $100\,\mathrm{km\,s^{-1}}$ respectively. The metallicity distributions are slightly different (at the $\approx 1.5\sigma$ level), the principal difference being an excess in (d) at [Fe/H] $\approx -1.25$. (e) is similar to (a), but only stars with $(B_J - R) > 1.8$ are displayed. The velocity distribution corresponding to (e) is shown in (f); note now the clear division into two populations. The metallicity distributions below and above the cut at $v = 100\,\mathrm{km\,s^{-1}}$ are shown in (g) and (h) respectively. These are significantly different (at the $\approx 3.5\sigma$ level) from each other.

(1990) has a mean heliocentric velocity of $v_{\mathrm{HEL}} = 22\,\mathrm{km\,s^{-1}}$, and therefore cannot be associated to the feature reported in the present paper ($v_{\mathrm{HEL}} = 140 \pm 2\,\mathrm{km\,s^{-1}}$). We used UKST $B_J$ and R plates to obtain a deep CMD emphasize that we have as yet no evidence to support or disprove this assumption. It is justified only in that it minimises the number of special explanations required here.

### 5.1 The Disk Hypothesis

Could the feature be part of a distant disk-like population? If we disregard those stars with $R > 18$ in the CMD of figure 5a, because we may be conservative about APM completeness problems near the plate limit, then the steeply rising arm from R = 17.7, $B_J - R = 1.4$ to R = 14, $B_J - R = 3.0$ could be interpreted as part of a subgiant branch of an old, metal rich cluster. The well populated region near R = 17.7, $B_J - R = 1.4$ would then be the base of the giant branch. This is depicted in figure 14, where we have superimposed three of the Revised Yale Isochrones (Green et al. 1987), shifted to $R - M_R = 14.5$ and corrected for extinction $A_R = 0.40$, on the photographic CMD of figure 5a. We tried to fit many such isochrones; no single curve fits both the excess near R = 17.7, $B_J - R = 1.4$ and the arm from R = 17.7, $B_J - R = 1.4$ to R = 14, $B_J - R = 3.0$.

There is no trace of the comoving group at negative longitudes either in the photometry or in the spectroscopic data. Thus, given that we know from the spectroscopy that we observed large numbers of Galactic bulge stars along the same lines of sight, we would require sufficient reddening to hide the comoving group, but not enough to hide the bulge. This could only be possible if the comoving group had a much shorter radial scale length than the bulge and if there were an opaque 'wall' of extinction near the tangent point to the line of sight at negative longitudes. In figure 15 we show the CMD from a $2.5° \times 2.5°$ degree scan at $\ell = -5°$, $b = -15°$, that is the mirror image reflection about the bulge minor axis of figure 5a. If these fields were identical, but for the reddening, then we crudely estimate that it would require more than one magnitude of extinction in excess of the value given in the maps of Burstein & Heiles to remove the narrow feature seen near R = 17.7 in figure 5a. Two globular clusters are known in the region near figure 15: NGC 6496 and NGC 6541; the extinction towards both clusters is in agreement with Burstein & Heiles values. However, since the clusters lie within 7 kpc of the Sun, they do not probe the total extinction down the line of sight. (Recall also that nearby, at $\ell = -10°$, $b = -10°$, Harding & Morrison (1993) find low and uniform reddening.)

If the extinction were in the foreground (within say $\approx 1$ kpc from the Sun), then the required one magnitude would be inconsistent with the observed difference in the luminosity functions, as we show in figure 16. However, we also show that a hypothetical opaque screen in their field, similar to those in figure 5, but this failed to show anything other than expected Galactic components.

Figure 13. The four carbon stars found in the $\ell = 5°$ fields. All have heliocentric radial velocities near $140 \, \mathrm{km \, s^{-1}}$.

### 4.2 Carbon Stars

Azzopardi et al. (1985) reported the first finding of carbon stars in the Galactic bulge; from photographic photometry they found $15.8 < V < 16.6$ and $2.0 < (B - V) < 2.3$. With their estimate of the extinction $A_V = 1.5$, $R_0 \approx 14.0$, $(B_J - R)_0 \approx 2.0$. These bulge carbon stars are $\approx 1/2$ magnitude brighter than those listed in table 1, but are generally about one magnitude bluer. Tyson & Rich (1991) find that the line of sight velocity dispersion of a large sample of bulge carbon stars is $\sigma_v \approx 113 \pm 14 \, \mathrm{km \, s^{-1}}$. These are kinematically very different from the four carbon stars listed in table 1, which clearly belong to the kinematically cold population of $\sigma_v \lesssim 15 \, \mathrm{km \, s^{-1}}$.

## 5 INTERPRETATIONS

We consider whether the comoving group could be part of the central regions of the Galactic disk, whether it is part of a distant spiral arm, or whether it belongs to a distant unknown dwarf galaxy. If this population is of Galactic origin (*i.e.* the first two hypotheses), its azimuthal velocity (about the Galactic centre) should be at most $\approx 250 \, \mathrm{km \, s^{-1}}$, so that the observed $172 \, \mathrm{km \, s^{-1}}$ (Galactocentric) mean radial velocity implies that the feature must be located within $\approx 1$ kpc from the tangent point to the line of sight; it would therefore be constrained to lie in the range $7 \, \mathrm{kpc} \lesssim d \lesssim 9 \, \mathrm{kpc}$, where $d$ is Heliocentric distance.

In the discussion below we will make the assumption that all of the unexpected colour-magnitude features (*cf* section 3.2) are related and part of the same object, and that the unexpected kinematic feature (*cf* section 3.1) samples the kinematics of that object. We of extinction near the tangent point to the line of sight in the $\ell = -5°$, $b = -15°$ field cannot be ruled out with these data alone.

Figure 14. This CMD is identical to figure 5a, but we have superimposed theoretical isochrones from Green et al. (1987) using $A_R = 0.35$ and $R - M_R = 14.5$. The isochrones have solar Helium abundance. The left isochrone has [Fe/H] = 0.3, $T = 2$ Gyr; the base of its giant branch fits the position of the observed concentration near R = 17.7, but the giant branch fit is poor. On the right we show an isochrone with [Fe/H] = 0.3, $T = 22$ Gyr, which gives a good giant branch fit, but does not explain the stellar overdensity near R = 17.7.

Figure 16. We show a comparison between the luminosity function (LF) created from the CMD in figure 5a (dots), figure 5b (filled squares) and a LF created from the CMD shown in figure 15 (star symbols). The extinction in these fields is clearly similar. Note the bump in the LF of figure 5a near R = 17.7. Certainly a difference in foreground extinction of the order of one magnitude can be ruled out. We have also plotted the expected LF (diamond symbols), according to our Galaxy model, where the integration down the line of sight was terminated at the tangent point, thus modeling the effect of a opaque screen of extinction. That scenario is still possible according to these data, but it is of course somewhat contrived.

Figure 15. This CMD is centred at $18^h 37.9^m$, $-40°5'$ (B1950) ($\ell = -4.6°$, $b = -15.3°$) and again covers a 2.5° × 2.5° area of sky. Qualitatively, it appears to be very similar to figure 5b.

To obtain some further constraints on the extinction, we investigate APM galaxy counts in the regions of sky corresponding to figure 5a and figure 15. The APM classification of galaxies is discussed at length in Maddox (1988). The analysis of regions at low Galactic latitude is extremely problematic with this technique, since the classification is accurate to 10%, while stars outnumber objects classified as galaxies by a factor of 10 down to the plate limit in these two fields, so we should expect at least 50% star contamination in the galaxy sample. Bearing this in mind, we present the corresponding galaxy LFs (corrected for extinction $A_R = 0.35$) in figure 17. The dashed line represents a 'by eye' fit to the R luminosity function compiled by Metcalfe et al. (1991). Because the galaxy counts are so greatly contaminated, no conclusions about the difference in the total line of sight extinction can be made from these data. However, since we find galaxies in both regions from visual inspection, the hypothetical opaque 'wall' of extinction in the $\ell = -5°$, $b = -15°$ field can be ruled out.

In summary, we assert that the extinction in the two regions is similar, that is $E_{B-V} = 0.14 \pm 0.03$.

In figure 18 we compare the luminosity function of the feature, as shown previously in figure 8, to a LF representative of the solar neighbourhood disk (Wielen 1974) which has been shifted to the distance of the tangent point to that line of sight. The two LFs are clearly very different from each other.

The relation that the carbon stars could have to the hypothesised disk is uncertain. Carbon stars are generally intermediate age objects, normally seen at the tip of the asymptotic giant branch (AGB). Yet the only isochrones that fit the proposed disk giant branch are extremely old. Younger isochrones not only fail to fit the giant branch, they also have too high a luminosity at the AGB tip. Alternatively, these carbon stars could be related to those studied in the central bulge

Figure 17. Showing the comparison between the galaxy LF created from selecting only APM classified galaxies from the same field as figure 5a (dots) with another galaxy LF created from the same field as figure 15 (star symbols). Extinction has been corrected for using $A_R = 0.35$. The dotted line is a fit to the R luminosity function compiled by Metcalfe et al. (1991) (corrected for area and bin size).

Figure 18. We show the comparison between the LF of the comoving group, to that of a solar neighbourhood disk population (Wielen 1974).

(the nature of which is not well known — *cf* Whitelock 1993), however their colour and kinematics are markedly different (*cf* section 4.2).

Another powerful objection to this hypothesis is that the velocity dispersion of the comoving group is $\approx 10$ times lower than the dispersion observed in Galactic old disk stars at similar longitudes by Lewis & Freeman (1989).

Assuming that the comoving group were part of the central disk, we investigate this scenario using Jeans' equations of hydrodynamic equilibrium. In cylindrical coordinates $(R, \phi, z)$ and assuming axisymmetry, the radial force equation can be written as

$$R\frac{\partial\Psi}{\partial R} - \overline{v}_\phi^2 = \sigma_{\phi\phi}^2 - \sigma_{RR}^2 - \frac{R}{\nu}\frac{\partial(\nu\sigma_{RR}^2)}{\partial R} - \frac{R}{\nu}\frac{\partial(\nu\sigma_{Rz}^2)}{\partial z}, (1)$$

where $\Psi$ is the potential of the stellar system, $\nu$ is the spatial density, $v_i$ is the mean velocity in the $e_i$ direction, and $\sigma_{ij}^2$ is a tensor describing the random velocities at a point.

The kinematic data we have gathered (*e.g.* figure 1) gives $\overline{v}_\phi$ and $\sigma_{\phi\phi}$ — via a model — both for the bulge and for the (assumed) disk population seen in the $5°$ fields. The remaining unknowns in equation 1 are the derivative of the potential, $\sigma_{RR}$, $\sigma_{Rz}$ and their derivatives. These quantities shall be examined individually.

In the plane of an axisymmetric galaxy $R(\partial\Psi/\partial R)_{R=0} = R^2\Omega_c^2$, where $\Omega_c$ is the angular circular velocity of the potential. This of course has been mapped with observations of HI emission (Burton 1988). However, finding $R(\partial\Psi/\partial R)$ at a point considerably out of the galactic plane (in the $5°$, $-20°$ field, the tangent point lies at $R \approx 0.7$ kpc, $z \approx 2.7$ kpc) requires unavailable knowledge of the mass distribution in the galaxy. Furthermore, the value of $\Omega_c$ derived from HI velocities near the galactic centre will be incorrect if the Milky Way is barred. However, if we have observed the inner disk as well as the bulge, we may use equation 1 on both populations to eliminate the term $\partial\Psi/\partial R$ in the three $\ell = 5°$ fields. In the $(\ell = -25°, b = -12°)$ field, $R > z$ and gas orbits are probably circular (Burton 1988), so we can use $\Omega_c$ (from HI) more confidently as a measure of $\partial\Psi/\partial R$.

The behaviour of $\sigma$ in a disk population can be estimated by assuming that the disk is self-gravitating and isothermal. Disks of external galaxies are observed to have constant thickness $h_z$ independent of $R$ (van der Kruit 1979) — though note that there is evidence that old disk stars are trapped in spiral arms (Pacziński et al. 1994). Then $h_z \propto \sigma_{zz}^2/\Sigma$, where $\Sigma \propto \exp(-R/h_R)$ is the disk surface mass density, so that $\sigma_{zz}^2 \propto \exp(-R/h_R)$. We make the common assumption that $\sigma_{RR}^2 \propto \sigma_{zz}^2$ in the disk components throughout the galaxy (*cf* van der Kruit 1990).

The behaviour of $\sigma$ in the bulge is assumed, then tested against the galaxy model.

To proceed, an assumption must be made regarding the direction along which the velocity dispersion tensor points. The simplest case is to assume that the velocity dispersion tensor $\sigma$ is diagonal in cylindrical coordinates, so that $\sigma_{Rz}^2 = 0$, $\sigma = \sigma(R, z)$ and $\overline{v}_\phi = \overline{v}_\phi(R, z)$.

We model the 'disk-like' component by a double exponential $\nu \propto \exp(-R/h_R - z/h_z)$, so that with the above assumptions equation 1 can be written as

$$R\frac{\partial\Psi}{\partial R} = \overline{v}_{d,\phi}^2 + \sigma_{d,RR}^2\left[\frac{\sigma_{d,\phi\phi}^2}{\sigma_{d,RR}^2} - 1 + \frac{2R}{h_R}\right], (2)$$

where the subscript $d$ has been used to denote disk-specific velocity and dispersion. Since the disk components in the $5°$ fields are observed at $\approx 170\,\mathrm{km\,s^{-1}}$ (Galactocentric), the stars must lie very close to the tangent point to those lines of sight, and the dispersion seen in the radial velocity distribution ($\lesssim 15\,\mathrm{km\,s^{-1}}$) is almost entirely $\sigma_{d,\phi\phi}$. We expect $\sigma_{d,RR}/\sigma_{d,\phi\phi}$ to be greater than 1, and less than $\approx 2$, since this is the range observed in all other galactic components (*cf* Gilmore 1990). With these assumptions, we find that the scale length $h_R$ of the comoving group must be less than the very unlikely value of $0.3\,\mathrm{kpc}$ (the old disk has $h_R = 3.5\,\mathrm{kpc}$) for the term in square brackets in equation 2 to make more than 10% of $\overline{v}^2_{d,\phi}$. We therefore make the approximation

$$R\frac{\partial\Psi}{\partial R} \approx \overline{v}^2_{d,\phi}, \qquad (3)$$

that is, the tangential velocity of the comoving group would balance the Galactic radial potential gradient to good approximation. Lewis & Freeman (1989) show that the central regions of the Galactic old disk population also has this property. However, the high radial potential gradient implied by equation 3 is at odds with a recent self-consistent bulge+disk model (Kent 1992) which fits essentially all central Galaxy stellar kinematics and also the kinematics of bulge stars in the fields of interest here, predicting $(R\partial\Psi/\partial R)^{1/2} \approx 70\,\mathrm{km\,s^{-1}}$ at $\ell = 5°$, $b = -15°$.

Stellar number density in the outer bulge is consistent with a power-law fall-off $(s/s_0)^{-3.7}$, where $s$ a flattened radial coordinate. Given the above assumption that the velocity dispersion tensor points along the $R, \phi, z$ axes, then equation 1 becomes

$$R\frac{\partial\Psi}{\partial R} = \overline{v}^2_{b,\phi} + \sigma^2_{b,RR}\left[\frac{\sigma^2_{b,\phi\phi}}{\sigma^2_{b,RR}} - 1 + 3.7 - \frac{R}{\sigma^2_{b,RR}}\frac{\partial\sigma^2_{b,RR}}{\partial R}\right]. (4)$$

From the bulge model, we find that our data are consistent with $\partial\sigma^2_{b,RR}/\partial R = 0$. Defining $\beta = \sigma_{b,\phi\phi}/\sigma_{b,RR}$, then

$$\beta = \sqrt{\frac{(\overline{v}^2_{d,\phi} - \overline{v}^2_{b,\phi})}{\sigma^2_{b,RR}} - 2.7} \qquad (5)$$

No galactic component has $\sigma_{\phi\phi} > \sigma_{RR}$, so $\beta$ must lie in the range $0 < \beta < 1$, which places the constraint

$$\sqrt{\frac{(\overline{v}^2_{d,\phi} - \overline{v}^2_{b,\phi})}{3.7}} < \sigma_{b,RR} < \sqrt{\frac{(\overline{v}^2_{d,\phi} - \overline{v}^2_{b,\phi})}{2.7}} \qquad (6)$$

giving $87 < \sigma_{b,RR} < 102\,\mathrm{km\,s^{-1}}$ in all the $5°$ fields. In the ($\ell = -25°$, $b = -12°$) region, we assume $R\partial\Psi/\partial R \approx R^2\Omega_c^2 = (215\,\mathrm{km\,s^{-1}})^2$ (Burton 1988), so that with $\overline{v}_{b,\phi} = 110\,\mathrm{km\,s^{-1}}$ at the tangent-point in that field, we find $98 < \sigma_{b,RR} < 112\,\mathrm{km\,s^{-1}}$.

However, if $\sigma_{b,RR} \approx 100\,\mathrm{km\,s^{-1}}$ in these fields, we find that the tangential bulge dispersion

$\sigma_{b,\phi\phi} \lesssim 40\,\mathrm{km\,s^{-1}}$ to get a plausible fit to the observed velocity distributions (*cf* figure 1). The resulting ratio $\sigma_R/\sigma_\phi \gtrsim 2$, stands in stark contrast to the inner regions of the bulge which display isotropic velocity dispersion (Spaenhauer et al. 1992).

In summary, there are clearly many strong objections to this interpretation, so we consider it very unlikely that the comoving group is connected with the inner Galactic disk.

### 5.2 The Spiral Arm Hypothesis

There is now some evidence that a significant fraction of old disk stars in the Milky Way are bound to spiral arms (Paczyński et al. 1994, Ramberg 1957). Such behaviour has also been recently observed in the spiral galaxy M51 (Rix & Rieke 1993).

The hypothesis that the unknown population is part of a spiral arm close to the Galactic centre solves some of the problems inherent in the above disk hypothesis. The narrow line of sight depth implied by the small magnitude spread of the excess feature near $R = 17.7$, and the low velocity dispersion is consistent with seeing a spiral arm approximately edge-on.

The only possible interpretation of the colour-magnitude relation is again that the excess population near $R = 17.7$ is the base of a giant branch, and the arm from $R = 17.7$, $B_J - R = 1.4$ to $R = 14$, $B_J - R = 3.0$ is a subgiant branch. The discussion in section 5.1 showed that no single plausible isochrone fits all of the CMD, and that the observed luminosity function (LF) is not consistent with a solar neighbourhood-like LF.

The non-axisymmetry of the feature is less problematic under this scenario, since spiral arms need not be axisymmetric (*e.g.* the Sb Galaxy NGC2841 *cf* Binney & Tremaine 1987), but note that some situations, such as a tightly wound spiral arm, or that depicted in figure 19 can be ruled out.

In summary, the carbon star problem remains unresolved, the metallicity estimates are in conflict with isochrones that could fit the giant branch, and the CMD and LF are inconsistent with plausible relations. Note also that since we observe the feature down to $b = -20°$, the spiral arms would have to be huge, extending out to more than $2.7\,\mathrm{kpc}$ below the Galactic plane.

### 5.3 The Dwarf Galaxy Hypothesis

An alternative explanation is that the feature is part of a hitherto unknown dwarf galaxy. Then the well populated excess at $R = 17.7$, $B_J - R = 1.4$ is a horizontal clump which extends blueward as a blue horizontal branch to $R \approx 18.3$, $B_J - R \approx 0.0$, and the rising

**Figure 19.** This sketch shows a spiral arm configuration that is incompatible with our data. The lines are drawn at (from top to bottom) $\ell = 5°$, $\ell = 0°$, $\ell = -5°$. The arrow shows the sense of Galactic rotation.

arm is a red giant branch.

If this hypothesis is correct, we can obtain an alternative estimate of the reddening by comparing the horizontal and giant branch morphology with a similar photographic CMD showing the intermediate age population of the Small Magellanic Cloud (SMC) (Gardiner & Hatzidimitriou 1992). By translating the SMC diagram 1.7 magnitudes brighter in R and 0.2 magnitudes redward in $B_J - R$, an excellent match ensues. The reddening estimated this way amounts to some 0.3 magnitudes of extinction in R, in good agreement with the previous value. Adopting this directly determined reddening, and assuming a distance to the SMC of 57 kpc (Feast 1991), leads to a preliminary distance estimate of 24±2 kpc for the Sagittarius dwarf. The good match with an SMC intermediate age stellar population CMD suggests that a significant intermediate age population is present in the Sagittarius dwarf, while the possibly significant blueward extension of the horizontal branch indicates that there may be also an underlying old stellar population present. This is seen in Galactic dwarf spheroidal satellites and implies that a significant population II variable star component should be present.

In figure 20 we investigate the range of cluster CMDs that could be associated with the feature if this hypothesis is correct. We find that the red arm of the dwarf galaxy (the upper regions of which was fit in figure 10) lies between the giant branches of the globular clusters 47Tuc ($[Fe/H] = -0.71$, Djorgovski 1993) and M5 ($[Fe/H] = -1.4$, Djorgovski 1993). A good fit to the giant branch is also obtained with a young (7 Gyr) isochrone of metallicity $[Fe/H] = -1.0$, though clearly many others between the extremes shown in figure 21 fit the CMD adequately well.

**Figure 20.** This CMD is identical to figure 5a, but fits to the CMDs of the globular clusters 47Tuc, M5 and M92 are superimposed. These are taken from Sandage (1982), and have been converted into ($B_J - R$), R (*cf* Ibata & Gilmore 1995); ($B_J - R$) colours are reddened by $E_{(B_J-R)} = 0.26$, R is made fainter by $A_V = 0.4$ and by $R - M_R = 16.9$. We also show the central line of the box drawn in figure 10 which was found to fit this region of the colour-magnitude feature to within the photometric error. Clearly the RGB lies between those of 47Tuc and and M5.

**Figure 21.** This CMD is identical to figure 5a, but we have superimposed theoretical isochrones from Green et al. (1987) using $A_R = 0.35$ and $R - M_R = 16.9$. The isochrones have solar Helium abundance. The left isochrone is for $[Fe/H] = -1.3$, $T = 25$ Gyr, the middle isochrone is for $[Fe/H] = -1.0$, $T = 7$ Gyr, while the isochrone on the right is for $[Fe/H] = -0.7$, $T = 6$ Gyr. Again, for reference, we have plotted the central line of the box drawn in figure 10 which was found to fit this feature.

Given the estimated 0.5dex error in the spectroscopically derived metallicities (*cf* figure 12h), we find that the Sagittarius dwarf has a metallicity spread statistically indistinguishable from that of Fornax (whose intrinsic metallicity spread has been found to be $\approx 0.3$dex — Sagar et al. 1990), as we show in figure 22. However, the spread in figure 12h may be in part due to contamination from the bulge at the high metallicity end and from the halo at the low metallicity end. Also, as noted in section 3.7, the metallicity values derived for the reddest stars in the sample are subject to considerable uncertainty because of the low number of very red standard metallicity stars available to determine the metallicity calibration; it is possible therefore that the random metallicity errors for these stars has been underestimated. However, if the intrinsic metallicity spread in Sagittarius is comparable to that in Fornax, the former should display an intrinsic giant branch colour spread similar to the $\approx 0.1^{\mathrm{m}}$ in $(B-V)$ observed in the Fornax giant branch (Sagar et al. 1990). Given the photometric *rms* error of $\approx 0.2^{\mathrm{m}}$, we can reject an intrinsic $\approx 0.1^{\mathrm{m}}$ giant branch spread at better than the $3\sigma$ level (*cf* figure 10). Conversely, if we require an intrinsic $\approx 0.1^{\mathrm{m}}$ giant branch spread, we need to have overestimated the photometric error by $\approx 0.07^{\mathrm{m}}$ (for rejection at less than the $2\sigma$ level).

The luminosity of such an extended system is difficult to estimate directly but by comparing the number of horizontal and giant branch stars with Galactic dSphs it is possible to make a provisional estimate. For example, Sagittarius has a similar central surface density of stars to that of the dSph Carina, but is roughly a factor of 10 larger, suggesting Sagittarius is of comparable luminosity to the dSph Fornax. The total number of horizontal branch stars within the outermost contour shown in figure 11 is $\approx 17000$. As stated in section 3.6, we estimate that that contour level traces approximately half of the mass of Sagittarius. The dwarf spheroidal Fornax contains $\approx 35000$ horizontal branch stars. With the caveat that the evolutionary history and hence detailed stellar mix may be different, the number of horizontal branch stars in Sagittarius and Fornax is also similar implying an absolute luminosity of $M_{\mathrm{V}} \approx -13$. If this dwarf galaxy follows the $M_{\mathrm{V}}$, [Fe/H] trend seen in dSphs (Armandroff et al. 1993), the above absolute luminosity suggests a mean metallicity value of $[Fe/H] \approx -1.4 \pm 0.3$, lower than the metallicity value ([Fe/H] $= -0.95$) estimated from the RGB colour. Conversely, if the Sagittarius dwarf has [Fe/H] $\approx -1$, its absolute luminosity would be $M_{\mathrm{V}} < -14$.

Therefore, if we are to maintain the dwarf galaxy hypothesis, we must either assume that the metallicity estimate (from the RGB colour) has a systematic error of $\approx 0.5$dex (though note that a $\approx 0.1^{\mathrm{m}}$ systematic error in $B_{\mathrm{J}} - R$ would be sufficient to give rise to that error), or we have to postulate that the Sagittarius dwarf is significantly more luminous than the largest currently known Galactic dSph, Fornax. The spectroscopic mean abundance is intermediate between that of the Small Magellanic Cloud (whose mean metallicity has been found to be [Fe/H] $\approx -0.5$, Feast 1991) and that of Fornax ([Fe/H] $\approx -1.5$, Sagar et al. 1990); this would imply a mass in the range $10^{9} \gtrsim M \gtrsim 10^{8}\,\mathrm{M}_{\odot}$.

Figure 22. We present the comparison between the spectroscopically derived metallicity distribution of the red, high velocity K giants (*cf* figure 12h) with a gaussian of $\sigma = 0.3$dex (which is the intrinsic metallicity spread of the Fornax dwarf — Sagar et al. 1990 — convolved with the estimated error distribution — a gaussian of $\sigma = 0.5$dex). The distributions are distinguishable at less than the $1.5\sigma$ level.

The colours of the carbon stars listed in table 1 are similar to the cool asymptotic giant branch stars found in Fornax and the Magellanic clouds. If these carbon stars are indeed AGB stars, their appearance would imply the presence of a significant intermediate age component. The only Galactic dSph with a large carbon star population is Fornax. To test this we examined a UKST objective prism plate, (with D.H. Morgan, private communication) and isolated candidate carbon stars from our APM colour-magnitude data. Spectra from SAAO, which will be published elsewhere, confirm that at least several tens of carbon stars are associated with the dwarf galaxy. That is, there is increasing evidence for a significant intermediate age AGB population.

In figure 23, we compare the LF of the Carina dwarf spheroidal (Mighell 1990) with the luminosity function of the feature (as shown in figure 8). $B_{\mathrm{J}} - R$ and $R$ have been converted to V, as described in Ibata & Gilmore (1995a), and we have corrected for the difference in extinction $A_{\mathrm{V}} = 0.47$, and for a difference in distance modulus $V - M_{\mathrm{V}} = 2.75$ (since Carina lies at $85 \pm 5$ kpc Irwin & Hatzidimitriou 1993, and the Sagittarius dwarf would lie at $24 \pm 2$ kpc). These two luminosity functions are in very good agreement with each other (though recall that the empirical LF was obtained by subtracting off a LF from a nearby patch off sky whose total counts had been normalised by reference to our Galaxy model; without this correction the fit can be rejected at the $2\sigma$ level). The Carina dSph has a strong intermediate age stellar component — the similarity of its LF with that of the proposed

Sagittarius dwarf galaxy is consistent with the latter also being dominated by an intermediate age population; this in turn is consistent with the discovery of red carbon stars in Sagittarius.

If the (weak) feature near $B_J - R = 0$, $R = 18$ can be confirmed to be a blue horizontal branch, then this, together with the existence of a carbon star population and a red clump, would imply that there is a significant age spread in the stellar population of this galaxy.

The tidal radius of a dwarf galaxy of mass similar to that of Fornax, at a position $\approx 15\,\mathrm{kpc}$ from the Galactic centre is $\approx 0.6\,\mathrm{kpc}$. However, stars belonging to the Sagittarius system are seen spread out over a radius of more than $3\,\mathrm{kpc}$, so it is being torn apart by the Milky Way's tidal force. The spatial structure of the object (section 3.6) is consistent with the dwarf galaxy hypothesis, while its irregular aspect is also consistent with ongoing tidal disruption. N-body simulations of satellite galaxies in the tidal field of the Galaxy (Oh et al. 1994) show that stars are pulled out along the direction of the orbit of the satellite. This would be consistent with figure 11, if the projection of the dwarf galaxy orbit points along the direction of its elongation.

Another consistency argument in favour of the dwarf galaxy hypothesis is the $\approx 10\,\mathrm{km\,s^{-1}}$ velocity dispersion, which is typical of dwarf satellite systems (*cf* Irwin & Hatzidimitriou 1993 for a summary of properties).

## 5.4 Globular Clusters

As noted before, the globular cluster M54 lies along the same line of sight as the densest part of the feature (its position is marked by a 'star' symbol in figure 11). We give some suggestive evidence that it may be linked to the object reported in this paper. M54 is listed as being at a distance of 21.2 kpc, consistent with the distance required by the dwarf spheroidal hypothesis. However, the colour-magnitude relation of M54 is different to that of the dwarf, as we show in figure 24, where we have superimposed a 'by eye' ridge-line fit to its CMD. With a metallicity listed as $[\mathrm{Fe}/\mathrm{H}] = -1.4$ (Djorgovski 1993), M54 is slightly more metal poor than the Sagittarius dwarf. A recent measurement of the heliocentric radial velocity of M54 gives $v = 140 \pm 2\,\mathrm{km\,s^{-1}}$ (Suntzeff, priv. comm.), identical to the velocity of the dwarf. The correlations both on the sky, in distance and in radial velocity firmly establish an association between the cluster M54 and the object reported in this paper.

A further three globular clusters in the region of sky near M54 lie at approximately the same distance and have approximately the same radial velocity, *cf* table 2. This coincidence is very suspicious: only $\sim 120$ Galactic globular clusters are known, yet these

Figure 23. The luminosity function of the comoving group (dots) is compared to that of the dwarf spheroidal galaxy Carina from Mighell (1990) (star symbols). The LFs are in good agreement (they are different at about the $1\sigma$ level).

Figure 24. This CMD was shown previously in figure 5a; we have superimposed a ridge-line fit to the CMD of M54 from Harris (1975). The short line shows the RGB fit of figure 10.

four lie in a very narrow spatial and kinematic range. In figure 25 we show all the Galactic globular clusters of known distance; the position of the outermost contour in the isopleth map of figure 11 is also shown. The four globular clusters listed in table 2 lie approximately along the direction of elongation of the dwarf galaxy. If these clusters are indeed associated with the dwarf galaxy, the latter would have to be significantly larger than the isopleth map (figure 11) suggests. This may be consistent with the high luminosity and large mass implied by the spectroscopic and RGB-derived metallicity estimates. Globular clusters are observed in the dwarf spheroidal galaxy Fornax, which possesses five, while the Large and Small Magellanic clouds possess eight and one respectively — Harris (1991), with these numbers depending sensitively on one's adopted definition of a globular cluster. Note that the metallicity of Ter 7 (*cf* table

2) is exceedingly high for a halo cluster; however, young ($< 2\,\mathrm{Gyr}$ old) clusters of similar metallicity are found in the SMC (Da Costa 1991). Note also that significant numbers of the dwarf galaxy K giants were found to have similar abundance to these globular clusters.

Table 2: The four globular clusters seen near $\ell = 5°$, $b = -15°$

| Cluster | $\ell_{II}$ | $b_{II}$ | $V_{HB}$ | $E_{B-V}$ | $v(\,\mathrm{km\,s^{-1}})$ | [Fe/H] |
|---|---|---|---|---|---|---|
| M54 | 5.608 | -14.086 | 17.71 | 0.14 | $140 \pm 2$ | $-1.42$ |
| Ter 7 | 3.387 | -20.066 | 17.80 | 0.06 | $162$ $157.6 \pm 3.1$ | $-0.49 \pm 0.09$ |
| Arp 2 | 8.547 | -20.785 | 18.21 | 0.11 | $119$ $120.0 \pm 2.6$ | $-1.73 \pm 0.03$ |
| Ter 8 | 5.759 | -24.561 | 17.85 | 0.12 | $132$ | |

$V_{HB}$ *is the apparent magnitude of the cluster horizontal branch; Arp 2 and M54 are from Webbink (1985), Ter 8 is from Ortolani & Gratton (1990), Ter 7 is from Armandroff, T., (priv. comm.).* $E_{B-V}$ *is the reddening in* $(B - V)$ *to the cluster and comes from Peterson (1993). All velocities listed in column 6 are Heliocentric radial velocities. The first row gives velocities for Ter 7, Arp 2 and Ter 8 from Armandroff, T., & Da Costa, G., (priv. comm.); M54 is from a recent measurement by Suntzeff (priv. comm.). The entries on the second row for Ter 7 and Arp 2 gives the velocity as measured by Suntzeff, N., (priv. comm.). The metallicity values of Ter 7 and Arp 2 are from Suntzeff, N., (priv. comm.) and that of M54 is from Djorgovski (1993).*

We imagine constructing a "box" around the Sagittarius galaxy in the form of a conical section with limits between $10\,\mathrm{kpc}$ and $20\,\mathrm{kpc}$ in Galactocentric radius and subtending $10°$ around $(\ell = 5, b = -15)$ on the sky. Assuming that the halo globular cluster population has approximately 100 members and that the number density falls off as $r^{-3.5}$ (Zinn 1985), then the expected mean density of globular clusters in the box would be $\bar{v} = 0.11$. Using Poisson statistics, we find that the probability that 4 clusters would randomly lie within the box is $P \approx 7 \times 10^{-6}$. However, since the position of M54 could be said to 'define' the centre of Sgr, we ask also the probability that 3 clusters would be found in the box; then $P \approx 2 \times 10^{-4}$. Thus it appears that these globular clusters are themselves significantly clustered and part of the dwarf galaxy.

It is interesting to note that if there are indeed $\approx 4$ globular clusters associated with this merging dwarf galaxy, only $\approx 20$ such mergers need to have happened in the history of the Milky Way to have produced the entire halo globular cluster system. Before assuming this happened however, one must also consider the age distribution of the clusters. The four clusters associated with the Sagittarius dwarf galaxy are not typical in their abundance distribution, and appear from preliminary evidence to have unusually young ages, relative to other globulars. Thus one is equally justified in drawing the opposite conclusion, that accretion of Sagittarius-like globular clusters cannot be a significant contributor to the field globular cluster system. Improved age and abundance data are equired before any valid conclusions either way can be drawn.

### 5.5 Satellite Orbits

We now obtain some constraints on the orbit of the dwarf galaxy. The primary information we have is that the dwarf galaxy has a heliocentric distance of $24\,\mathrm{kpc}$ and a (Galactocentric) radial velocity of $170\,\mathrm{km\,s^{-1}}$ in the field at $(\ell = 5°, b = -15°)$, and that the radial velocity between $b = -12°$ to $b = -20°$ varies by less than $\approx 5\,\mathrm{km\,s^{-1}}$. We assume that at its current position, its motion is dominated by a logarithmic potential

$$\Psi = \frac{1}{2} v_0{}^2 \ln\left(R^2 + \frac{z^2}{c^2}\right), \tag{7}$$

where $v_0$ $(= 220\,\mathrm{km\,s^{-1}})$ is the circular velocity of the Galactic potential and $c$ is a flattening ratio. We also assume that the rotation of the dwarf galaxy contributes negligibly to the mean velocity in each field. Orbits that have the above property are found by integrating the equations of motion in this potential. There are essentially two possible orbits whose exact paths are slightly dependent on the flattening ratio $c$: for $c = 0.6$ the total velocity in the $(\ell = 5°, b = -12°)$ field would be $v = 310\,\mathrm{km\,s^{-1}}$, while if $c = 1.0$ then $v = 325\,\mathrm{km\,s^{-1}}$. (The latter implies a proper motion of $2.4\,\mathrm{mas/yr}$). The orbits have to be 'concave' with respect to the Sun, so as to produce the small $< 5\,\mathrm{km\,s^{-1}}$ difference in mean velocity between the fields. A sketch of this is given in figure 26, where we show the two possible orbits under the above assumptions. If the dwarf galaxy is on the orbit that is moving away from the Galactic disk, then its last impact with the disk occurred $\approx 1.5 \times 10^7\,\mathrm{yr}$ ago, at $\approx 15\,\mathrm{kpc}$ from the Galactic centre.

The strong tidal disruption that this dwarf galaxy is currently suffering suggests that it would not be able to survive many orbits — inverting this argument suggests that it is probably on its first pass or one of it first passes near the Galaxy. So we would expect to find it on an orbit with period of order a Hubble time. The perigalactic velocity of such an orbit depends on the extent of the dark halo. Assuming the dark halo extends out to $\approx 100$ kpc with constant circular velocity $v_0 = 220$ km s$^{-1}$, and that teh Sagittarius dwarf is on an orbit with perigalactic distance $\approx 10$ kpc, then an implied apogalactic distance of $\approx 350$ kpc gives an orbital period $10^{10}$ years, and an perigalactic velocity of $\approx 400$ km s$^{-1}$. A smaller perigalactic velocity ($\approx 350$ km s$^{-1}$) is possible if the dwarf is currently at its perigalactic distance ($\approx 15$ kpc).

Figure 25. We show a zenith projection of the entire sky; the lines are spaced at intervals of 20 degrees in Galactic longitude and latitude. The centre is at ($\ell = 0°, b = 0°$), and galactic longitude increases to the left. The lowest contour of the excess population previously plotted in figure 11 can be seen in the region near ($\ell = 5°, b = -15°$). All Galactic globular clusters of known distance have also been plotted. Dots represent those with Galactocentric distance of less than 12.5 kpc, filled circles represent those between 12.5 kpc and 22.5 kpc, and star symbols represent those beyond 22.5 kpc. Note the four globular clusters with Galactocentric distance $\approx$ 16 kpc (filled circles) in the neighbourhood of the dwarf galaxy; these also have velocities similar to that of the dwarf (*cf* table 2).

The effect that merging satellites have on their host galaxies is currently an active field of research. Satellites (of $\approx 10\%$ the mass of the disk) tend to disrupt galactic disks, significantly changing their morphology and creating starburst galaxies (Mihos & Hernquist 1994). A primary constraint on the merger history of the Milky Way is the fragility of the thin disk, which requires that there has been no major merging event since the formation of the thick disk. However, merging with large numbers of small satellites is allowed, as long as the total mass of accreted satellites in the last 5 Gyr is below $\approx 4\%$ of the mass inside the solar circle (Tóth & Ostriker 1992). The merger with a satellite of mass $\approx 0.1\%$ of the mass of the disk

(Fornax has $\approx 0.5\%$) would have virtually no effect on the global properties of the Galaxy (Mihos, C., private communication).

### 5.6 Shocked Inter-Stellar Medium

If the dwarf galaxy has or had a collisional component (ie. gas or dust), it can be expected to interact with Galactic gas. We now calculate the kinematic signature of such an interaction. The state of Galactic gas in the region currently inhabited by the dwarf galaxy is not well known. Several lines of evidence indicate the presence of a halo of gas reaching out to a few tens of kpc, with this gas being a mixture of different temperatures from $\approx 10^4$K to $\approx 10^6$K (see *e.g.* Savage 1987 for a brief review). The $v_s \approx 400$ km s$^{-1}$ shock velocity of the dwarf will be supersonic even to the hottest ($\approx 10^6$ K) gas. To find whether the shock is adiabatic, we investigate the ratio of the cooling time of gas in the shock to the dynamical time of the shock front passing through that gas. Following the treatment for ionised gas from Lang (1980), we find $t_{\rm Cool}/t_{\rm Dyn} \approx 2.3 \times 10^{12} T^{-5/2} \ln \Lambda$, where $\Lambda \approx 1.3 \times 10^4 T^{3/2} N_e^{-1/2}$ and $N_e^{-1/2}$ is the electron density. If we use the rough estimate

Figure 26. This shows a sketch of the present position and velocity of the dwarf galaxy. The 'star' symbol marks the position of the Sun, the ellipse represents the Galactic Bulge, the filled circle is the Galactic centre, the observed lines of sight are represented with dotted lines, and the vectors show the direction of the orbits. Though the Sun is unlikely to be in orbital plane of the dwarf galaxy, it is very approximately opposite the Galactic centre to the dwarf at the current time, allowing us to sketch the Sun, the Galactic centre and the dwarf galaxy in the same plane. That plane is inclined by $\approx 20°$ to the plane perpendicular to the Galaxy containing the Sun and the Galactic centre.

$n(\mathrm{H}^+) \approx 10^4$ atoms cm$^{-3}$ (Savage 1987) (this term appears only in the logarithm, so its value is relatively unimportant), then the shock is adiabatic for gas of temperature below $\approx 3 \times 10^5$ K. In the case of a highly supersonic adiabatic shock, it is easy to show (Lang 1980) that the post shock velocity $v_b$ of a gas that has been shocked by a wave of velocity $v$ is

$$v_b = \frac{2}{\gamma + 1} v, \qquad (8)$$

where $\gamma$ is the usual ratio of specific heats. Thus with $\gamma = 5/3$, the (heliocentric) line of sight velocity of the post shock gas is $115\,\mathrm{km\,s}^{-1}$. This velocity should be observed in H$\alpha$ if the column density of Galactic gas is sufficiently large. However, no obvious H$\alpha$ emission could be seen from visual inspection of R-band UKST plates in this region. It is interesting to note that an asymmetric feature with heliocentric velocity $\approx 120\,\mathrm{km\,s}^{-1}$ can be seen in the Leiden $-$ Green Bank HI survey (Burton 1985) survey near $\ell = 5°$ and covering a latitude range $-19 < b < -11$, similar to that of the dwarf. However, any neutral hydrogen would be rapidly ionised by the massively supersonic shock, so the shocked gas is unlikely to emit in HI.

5.7 **Summary** In summary, the dwarf galaxy hypothesis explains the discovery of an unexpectedly large number of carbon stars, the mean metallicity from the spectroscopy is consistent with that derived from the colour of the giant branch, all of the features of teh stellar colour-magnitude relation which are not accounted for by teh Galactic foreground can be explained with a single object, the observed velocity dispersion and the spatial structure fit the hypothesis, and the reddening derived by comparison of its colour-magnitude diagram with that of the Small Magellanic Cloud is in good agreement with values from Burstein & Heiles. However, the spread of the spectroscopically derived metallicity distribution is not consistent with the colour spread of the red giant branch, and the mean metallicity estimates indicate that, if this dwarf galaxy is of the spheroidal class, it would have to be special in being the most massive known, possibly half way in mass between Fornax and the SMC, a dwarf irregular. We emphasise that the evidence to support the dwarf spheroidal, as opposed to a dwarf irregular, classification is at present very limited. Either is possible.

6 **CONCLUSIONS**

A spectroscopic data set obtained by Ibata & Gilmore (1995a,1995b) for the study of bulge K giants unexpectedly revealed a group of $\approx 100$ stars whose anomalous kinematics (mean heliocentric velocity $140\,\mathrm{km\,s^{-1}}$, velocity dispersion $\approx 10\,\mathrm{km\,s^{-1}}$) cannot be accounted for by Galaxy models. We obtain a photographic colour-magnitude diagram (CMD) along the same line of sight which reveals a large excess of stars ($\approx 10^5$ stars down to $R = 18$ in a $1.5^\circ \times 1.5^\circ$ field). Upon mapping this excess population over the sky, we find that it extends over an area of about $15^\circ \times 5^\circ$. A preliminary luminosity function (LF) for the object is also derived.

We investigated whether this group of stars could be intrinsically Galactic, but found several problems with this interpretation. On balance from teh available evidence of this paper, we conclude that the dwarf galaxy hypothesis gives the most plausible explanation for all the observed phenomena.

From the estimate of the dwarf galaxy's mass ($10^8 \lesssim M \lesssim 10^9\,\mathrm{M_\odot}$) and from its Galactocentric distance ($24 \pm 2\,\mathrm{kpc}$), we deduce that it extends to almost an order of magnitude further than its tidal radius. A large proportion of its stars are therefore being stripped, and are being merged into the halo of the Milky Way. This conclusion is somewhat dependent on the assumed mass and distance; the distance to the dwarf is moderately well determined (given the uncertainties in reddening and photometry), however its mass is currently only poorly known. To determine the rate of disruption, a better estimate of the mass of the object will be necessary.

Four globular clusters lie in approximately the same phase space region as the dwarf galaxy. We show that it is extremely likely that at least some of the four clusters are associated with the stellar population we have discovered. This is a very important result, proving that at least some halo globular clusters, as well as at least some halo field stars, originate from recent merger events.

Further investigation is motivated by the hypothesis that the merging of dwarf galaxies onto their larger companions may be a crucial factor determining the chemical, morphological and kinematic evolution of large galaxies like the Milky Way. The study of a nearby example of such an interaction provides an ideal opportunity to quantify this proposition.